%Paper: hep-ph/9310313
%From: sterman@nuclear.physics.sunysb.edu
%Date: Tue, 19 Oct 1993 11:22:12 EST

\input phyzzm
\rightline {ITP--SB--93--61}
\rightline{19 October 1993}

\title{Principal Value Resummation}

\author{Harry Contopanagos$^{1,2}$ and George Sterman$^1$}
\address{$^1$Institute for Theoretical Physics, SUNY Stony Brook\nextline
Stony Brook, New York 11794-3840}
\address{$^2$High Energy Physics
Division, Argonne National Laboratory\nextline
Argonne, IL 60439-4815}

\abstract
We present a new resummation formula for the Drell-Yan cross section.
The formal resummation of threshold corrections
in Drell-Yan hard-scattering functions
 produces an exponent with singularities from the
infrared pole of
the QCD running coupling. Our reformulation
treats such `infrared renormalons' by
a principal value prescription, analogous to a modified
Borel transform. The
resulting expression
includes all large
threshold corrections to the hard scattering function as an
asymptotic series in $\alpha_s$, but is a
finite function of $Q^2$.  We find that the ambiguities
of the resummed perturbation theory imply
the presence of higher
twist corrections to
quark-antiquark hard-scattering functions that begin at $\Lambda_{QCD}/Q$.
This suggests an important role for higher twist in the
phenomonolgy of hadron-hadron inclusive cross sections.
We also discuss the numerical evaluation of the exponent
and its
asymptotic perturbation series
for representative values of $Q^2$.
\endpage

\chapter{Introduction}

It has been known for some time that perturbative QCD corrections
to the inclusive Drell-Yan and other hadron-hadron
hard scattering cross sections are numerically important,
even at moderately high energies\Ref\rone{G.\ Altarelli, R.K.\ Ellis and
G.\ Martinelli, Nucl.\ Phys.\ B157 (1979) 461; B.\ Humpert and W.L.\ van
Neerven, Phys.\ Lett.\
84B (1979) 327; J.\ Kubar-Andr\'e and F.E.\ Paige, Phys.\ Rev.\ D19 (1979) 221;
K.\ Harada, T.\ Kaneko and N.\ Sakai, Nucl.\ Phys.\ B155 (1979) 169;
B165 (1980) 545 (E); T.\ Matsuura,
S.C.\ van der Mark and W.L.\ van Neerven, Phys.\ Lett.\ 211B (1988) 171; Nucl.\
Phys.\ B319
(1989) 570.}. Considerable effort
has been devoted to developing all-order resummation techniques in order
both to study the convergence of the QCD perturbation series and to control
the size of perturbative corrections to these processes.
In a recent paper\Ref\rtwo{H.\ Contopanagos and G.\ Sterman, Nucl.\ Phys.\ B400
(1993) 211.}, we presented an analysis of the resummed
threshold corrections
in the Drell-Yan cross section, deriving an explicit expression for
the hard scattering function directly in momentum space. This new resummation
formula organizes
all large and order unity threshold corrections
in the entire region where perturbation theory gives the dominant contribution.
Because the resummed expression involves integrals over
the scales of running couplings, it is
undefined when
these scales reach $\Lambda_{QCD}$.
The purpose of the present work is to address these and
related issues.

Beyond specific applications to dilepton cross sections,
we are concerned here with some very general problems in the application
of perturbative QCD to hard processes.  A complete picture of such
processes requires the inclusion of power-suppressed, or `higher twist'
contributions, but to include such contributions, we must first
define the `leading-twist', or perturbative series
\Ref\rthree{A.H.\ Mueller, Nucl.\ Phys.\
B250 (1985) 327; F.V.\ Tkachov, Phys.\ Lett.\ B125 (1983) 85.
}\Ref\rfour{A.H.\ Mueller in proceedings
of {\it QCD - 20 years Later}, Aachen, 1992
(to be published), Columbia preprint CU-TP-573 (1992);
Columbia preprint CU-TP-585 (1993).}.  To do so fully
is a formidable task, beyond our present abilities.  The class of large
perturbative threshold corrections to dilepton
(and other) inclusive hard scattering cross sections, however, is known
to all orders.  It is, as expected, ambiguous.
Our goal, then, will be to `make room' for higher twist by
defining the resummation of large corrections at leading twist.
At the same time, we may also ask why, despite all these
large corrections and the ambiguities in perturbation theory,
perturbative corrections at the one loop level are not totally
wrong.  While we cannot fully answer this question either,
we will get some hints toward its resolution.

The situation for dilepton production
hard-scattering functions is analogous, but not identical, to the
situation for the $e^+e^-$ annihilation total cross
section\refmark\rthree\refmark\rfour. In
$e^+e^-$, the ambiguities of perturbation theory
may be identified with the gluon condensate $\langle 0|F^2|0 \rangle$,
which appears in the operator product expansion for this
process.  Such contributions are
suppressed by the power $Q^{-4}$ relative to fixed-order perturbation
theory.  In contrast, we shall identify below the first ambiguity
in the resummed perturbative cross section at the much larger
level of $Q^{-1}$ relative to leading power.
This suggests an important phenomonological role for higher
twist in hadron-hadron scattering, at least at moderate energies.
Another difference between
dilepton production and the $e^+e^-$ annihilation cross section is
that in the latter case
low order corrections are much smaller. It is possible that
the occurrence of nonperturbative corrections at lower twist
in hadron-hadron scattering
is related to the larger size of its perturbative corrections.

Let us now review the basic results of the resummation program
applied to the dilepton cross section \refmark\rone\refmark\rtwo
$$h_1(p_1)+h_2(p_2)\to l\bar{l}(Q^\mu)+X\, ,
\eqn\twoone
$$
with a produced lepton pair of momentum $Q^\mu$.
In perturbation theory, the factorized form of this cross section is given by
$${d\sigma\over dQ^2}=\sigma_0\sum_{ab}\int_0^1{dx_a\over x_a}{dx_b\over x_b}
\phi_{a/h_1}(x_a,Q^2)\phi_{b/h_2}
(x_b,Q^2)
\omega_{ab}(\tau/x_ax_b,a_s(Q^2))\, ,
\eqn\twotwo
$$
where
$\sigma_0$ is the
Born cross section,
$\phi_{a/h_1}(x_a,Q^2)$ is the distribution function of parton $a$ in hadron
 $h_1$, and $\omega_{ab}$ is a short-distance function, or `hard part'.
In eq.~\twotwo\ we denote $\tau=Q^2/s$, with $s=(p_1+
p_2)^2$.  We emphasize that when we speak of `leading' and
`higher twist' in this paper, we are referring to the hard parts,
and not to the cross section as a whole, which depends on the
interplay of the short-distance functions with the evolving
distributions.

In eq.~\twotwo\ the hard parts $\omega_{ab}$, calculable in perturbation
theory, as well as the
non-perturbative parton densities, are not unique. We shall identify
 the distribution
 $\phi_{a/h_i}$ with the contribution, $F_{a/h_i}$,
  of parton $a$ to a structure function
in deeply inelastic scattering of a nucleon $h_i$.
Then the hard parts $\omega_{ab}$, which directly give a
 prediction
for the normalization of the Drell-Yan cross section in terms of the
observables
$F_{a/h_i}$,
 contain large corrections. These  come from large numerical
coefficients
 of $\delta$-functions, and from `plus distributions'.
The organization of such large corrections to all orders in perturbation
theory, is the
 content of
various resummation techniques\Ref\rgs{G.\ Sterman, Nucl.\ Phys.\ B281 (1987)
310.}\Ref\rfive{D.\ Appell, P.\ Mackenzie and G.\ Sterman,
Nucl.\ Phys.\ B309 (1988) 259.}\Ref\rcattren{S.\ Catani and
L.\ Trentadue, Nucl.\ Phys.\ B327 (1989) 323; B353 (1991) 183.}\Ref\rmagst{L.\
Magnea and G.\ Sterman, Phys.\ Rev.\ D42
(1990) 4222.}.
It is convenient, in order to study the large
perturbative
contributions in eq.~\twotwo\ which come from the region $z=\tau/x_ax_b\to1$ of
the integrals, to take moments
and study the large-$n$ region.
The only quantities that are singular in the $z\to 1$ limit
are the diagonal-flavor quark-antiquark
hard parts, for which we obtain
the exponentiated form\refmark\rtwo
$$
\tilde{\omega}_{q\bar{q}}(n,Q^2)=e_q^2 A(\alpha_s(Q^2))\tilde{I}(n,Q^2)\, ,
\eqn\twonine
$$
where the terms that are singular in the limit $n\to \infty$ are contained in
$$\eqalign{
\tilde{I}(n,Q^2)&={\rm exp}\Biggl[-\int_0^1dx\biggl({x^{n-1}-1\over
 1-x}\biggr)
\Biggl\{\int_0^x{dy\over 1-y}
g_1(\alpha_s[(1-x)(1-y)Q^2])\hfill\cr
&\hbox{\hskip 3.0 true cm} +g_2(\alpha_s[(1-x)Q^2])\Biggr\}\Biggr]
\cr
\hbox{\hskip 1.0 true cm} &\equiv \exp[E(n,Q^2)]\ .
}
\eqn\twoten
$$
In eq.~\twonine, $A(\alpha_s(Q^2))$
represents $\delta(1-z)$ contributions, including
 the exponentiated Sudakov
$\pi^2$ terms \refmark\rmagst.
$\tilde{I}(n,Q^2)$ contains the exponentiation of all plus-distributions, which
 are the source of
growth with $n$.
In ref.~\rtwo,  we employed an asymptotic expansion for
the exponent of eq.~\twoten,
to obtain, directly in momentum
space, the resummed hard parts in the form
$$
I(z,\alpha_s)=\delta(1-z)-\Biggl[{{\rm e}^{E\left({1\over 1-z},
\alpha_s\right)}\over \pi(1-z)}
\Gamma\biggl(1+P_1\biggl({1\over 1-z},\alpha_s\biggr)\biggr){\rm sin}\biggl(\pi
P_1\biggl({1\over 1-z},\alpha_s\biggr)\biggr)\Biggr]_+\, ,
\eqn\additionb
$$
where
$$P_1(n,\alpha_s)
\equiv
{\partial\over \partial\ln n}E(n,\alpha_s)\ .
\eqn\additionc$$
The central issue that we did not address in ref.~\rtwo\  was the
nature of this asymptotic approximation and its relation to non-perturbative
effects. This is the main subject of the present work.

Even though $\omega_{q\bar{q}}$ is finite at any finite order of perturbation
theory, the exponentiated
form, eqs.~\twoten, \additionb, suffers from
singularities associated with the behavior of the running coupling
 at small energy scales.
Therefore, even though  the above resummation formulas
are useful in reproducing finite-order results\Ref\rsix{L.\ Magnea, Nucl.\
 Phys.\
B349 (1991) 703.},
a direct comparison with phenomenology can be made only if a regularization of
soft-gluon effects is
supplied. Hence, a sensitivity of the resummed predictions to
the treatment of this `soft' region, $x\to 1$ in eq.~\twoten,
 is to be expected\refmark\rfive .
 For an infrared safe function, such regions give
contributions that are finite order-by-order in perturbation theory,
but which diverge in the sum to all orders.
This behavior is, in fact,  ubiquitous in QCD
\refmark\rthree\Ref\rthooft{'t Hooft, in {\it The Whys
of Subnuclear Physics,
Erice 1977}, ed.\ A.\ Zichichi (Plenum, New York 1979);
Y.\ Frishman and A.R.\ White, Nucl.\ Phys.\ B158 (1979) 221;
J.C.\ Le Guillou and J.\ Zinn-Justin, ed., Current
Physics -- Sources and Comments, Vol.\ 7,
{\it Large-order behaviour in perturbation theory}
(North Holland, Amsterdam, 1990);
G.B.\ West, Phys.\ Rev.\ Lett.\ 67 (1991) 1388;
L.S.\ Brown and L.G.\ Yaffe, Phys.\ Rev.\ D45 (1992) R398;
L.S\ Brown, L.G.\ Yaffe and C.\ Zhai, Phys.\ Rev.\ D46 (1992) 4712;
V.I.\ Zakharov, Nucl.\ Phys.\ B385 (1992);
M.\ Beneke and V.I.\  Zakharov,
Max Planck Inst.\ Munich preprint MPI-PH-92/53;
M.\ Beneke, Max Planck Inst.\ Munich preprint MPI-PH-93/6.
}.
In addition to its applications to the Drell-Yan and related
cross sections, the resummation \additionb\ may thus serve as a
`laboratory' to test our ideas about the
asymptotic nature of the QCD perturbation series.

In this paper, we shall propose a finite definition of the
exponent $E(n,Q^2)$ in eq.~\twoten\ as a {\it principal value}
integral in the moment variable $x$.
In this fashion we will define unambiguously - albeit not uniquely -
 the  perturbative content
of large corrections to the hard scattering function.  We will
see as well that any such definition requires that nonperturbative
corrections to the
hard scattering functions  begin at order $Q^{-1}$, a full
power of momentum transfer larger than in spin-averaged
deeply inelastic scattering, and three powers larger than in
the total $e^+e^-$ annihilation cross section.

Let us  reemphasize that our choice for a principal value resummation
is not unique.  The choice to be given below is
closely linked to the specific form for
the resummed hard part given in eq.~\twoten.  Other expressions for
$\omega_{q{\bar q}}$, which
sum the same sets of large perturbative contributions, but which differ in
nonleading terms, are possible.  For such expressions, a different use
of the principal value, or even the use of some other prescription, may be
advantageous.  Differences in these prescriptions will be reflected in
differences in higher twist corrections\refmark\rthree\refmark\rfour.
In fact,
there is a whole {\it class} of
resummation formulas, which can be constructed by exploiting the
analytic structure of the perturbative running coupling.
In addition to studying our
specific construction, we would like to argue
 that these issues bear further investigation
in hadron-hadron scattering cross sections.

In section 2 we define our principal value resummation formula for
$\omega_{q{\bar q}}(n,Q^2)$, and discuss its
 relation to the
Borel transform.  We relate the exponentiated
principal value prescription to
an analogous prescription for the evolution
equation satisfied by $\omega_{q{\bar q}}(z,Q^2)$.
We go on to evaluate analytically the leading (one-loop in the $g_i$
and the running coupling)
exponent in eq.~\twoten.
We then discuss  the consequences
of the principal value prescription for higher twist.  Here
we identify contributions at order $1/Q$,
associated with the ambiguities of the resummed perturbation theory.
In section 3, we
construct an explicit asymptotic perturbation series
for the exponent, and in section 4
 we explore the exponent and its
 asymptotic series numerically, again approximating it
 by the one-loop terms of the functions $g_i$.
 We find that a truncated perturbative series can give an excellent
 approximation to the exponent over a wide range in the variable $n$,
 but that such approximations require an energy-dependent
 maximum order, which differs for different terms in $E(n,Q^2)$.
 We shall also see that the full principal value exponent
 increases to a moderate value as $n$ increases, and then
 turns over and decreases as $n\rightarrow \infty$. Finally, in section 5 we
summarize our results and conclusions,
and discuss the outlook for applying this method to the calculation
of other hard hadronic processes.

\chapter{The Principal Value Exponent}

As explained in the introduction, we will construct a resummation formula
 analogous to
eq.~\twoten, that is calculable without reference to explicit IR cutoffs.
We shall deal directly
with the exponent $E(n,Q^2)$
given by \twoten. This defines the resummation
in both moment and momentum space.

\section{Definition}

Denoting the expression in the curly brackets of eq.~\twoten\ by
$$
\Gamma_{q{\bar q}}(1-x,Q^2)\equiv \int_0^x{dy\over 1-y}g_1(\alpha_s[(1-x)(1-y)
Q^2])
+g_2(\alpha_s[(1-x)Q^2])\, ,
\eqn\twotwelve
$$
we {\it define} the principal value exponent for the quark cross
 section
in moment space through
$$
\eqalign{
E(n,Q^2) &= -\int_P d\zeta
\biggl({\zeta^{n-1}
-1 \over 1-\zeta}\biggr)\Gamma_{q{\bar q}}(1-\zeta,Q^2)\
\cr
&\equiv -{1\over 2} \bigg \{
\int_{0+i\epsilon}^{1+i\epsilon} d\zeta
\biggl({\zeta^{n-1}
-1 \over 1-\zeta}\biggr)\Gamma_{q{\bar q}}(1-\zeta,Q^2)
+
\int_{0-i\epsilon}^{1-i\epsilon} d\zeta
\biggl({\zeta^{n-1}
-1 \over 1-\zeta}\biggr)\Gamma_{q{\bar q}}(1-\zeta,Q^2)
\bigg \}\, .}
\eqn\twothirteen
$$
Wherever $\Gamma_{q{\bar q}}(1-\zeta,Q^2)$ is an analytic function between
$0$ and $1$, we can deform the two contours
back to the real axis.
This is the case  at any finite order in perturbation theory.
The resummed exponent, on the other hand,  contains
IR divergences near $x=1$
{\it for any value of n} from the running of the coupling.
 In this region, the
quark cross section defined through $E(n,Q^2)$
remains well defined with the principal value prescription, although,
of course, contributions from this region are still infrared-sensitive.

How may we motivate the prinicpal value prescription?  In some
sense, the choice is quite arbitrary.  But some choice is necessary
if we are to `make room' for higher-twist effects in physical situations
where, as in dileption cross sections, perturbative corrections are
large.  In our opinion, it is precisely such quantities from which we may
eventually learn the most about the interplay of perturbative and
nonperturbative effects in QCD.
With these observations in mind, we may offer a
formulation of principal value resummation that is somewhat
more general than the specific integrals in eq.~\twoten.

The principal
value prescription may be used to
define a resummed series {\it whenever} the summation
of a set of perturbative contributions can be expressed as
an integral over the running coupling.
Infrared renormalons\refmark\rthooft are the simplest examples of this form.
  This set of contributions is
identified by a behavior from individual
diagrams at $n$th order of $ \alpha_s^n(Q^2) b_2^n n!$,
with $Q^2$ any
fixed scale.  Such a series of terms may be
generated by expanding the running coupling
$\alpha_s(k^2)$ inside the simple integral
$$
I(\alpha_s(Q^2))  \equiv \int_0^{Q^2} d k^2 k^2 \alpha_s(k^2)
\eqn\Ialphas
$$
as a power series in $\alpha_s(Q^2)$, using the one-loop expression,
$$
\alpha_s(k^2) =
{\alpha_s(Q^2) \over 1+(b_2/\pi)\alpha_s(Q^2)\ln(k^2/Q^2)}
={\pi \over b_2 \ln(k^2/\Lambda^2)}\, .
\eqn\runningalpha
$$

Many of the properties of infrared renormalons are brought
out very clearly by reexpressing the series as a Borel transform,
$$
\tilde{I}(b) = {1 \over 2\pi i}\int_{-i\infty}^{i\infty}
d(1/\alpha_s(Q^2)) e^{-ib/\alpha_s(Q^2)}
I(\alpha_s(Q^2))\, ,
\eqn\boreltrans
$$
whose formal inverse is
$$
I(\alpha_s(Q^2))
=
\int_0^\infty db\, e^{-b/\alpha_s(Q^2)} \tilde{I}(b)\, .
\eqn\invborel
$$
The integral $I(\alpha_s(Q^2))$ is undefined
as it stands, because of the singularity in the
running coupling $\alpha_s(k^2)$
 at $k^2=\Lambda^2$.  This singularity
is directly reflected in the $n!$ behavior of the
expansion in $\alpha_s(Q^2)$, and in a consequent
singularity on the positive real axis in
the Borel transform, $\tilde{I}(b)$, at $b=2\pi/b_2$.
In Ref.~\rthree, it was shown  that, given
what we know now about the perturbative series,  this
is the leftmost singularity of the transform.  Assuming the
inverse transform, eq.~\invborel, has been defined,
such a singularity may be expected to contribute only
at the level $Q^{-4}$ to the cross section, corresponding
exactly to the contribution of the gluon condensate,
$\langle 0|\; F^{\mu\nu}F_{\mu\nu}\; |0\rangle$.
One way to define the perturbative, and hence nonperturbative,
contribution at this level is to define the inverse Borel
transform as a principal value \refmark\rthree.
In fact, for the one-loop
running coupling the resulting expression is {\it exactly}
what we find by treating the original
integral
in eq.~\Ialphas\ as a principal value in $k^2$, and by
changing
 variables to
$$
b'\equiv \ln (k^2/Q^2)\, .
\eqn\bprimedef
$$
A similar change allows us to reproduce two-loop results
as well.

In the somewhat more complex situation of eq.~\twoten,
integrals of the running coupling appear in an exponent,
and the relation of a principal value
in the moment integrals to an inverse Borel transform
for $I(n,Q^2)$ is
not quite so simple.   The prescription can still be given a
reasonably general
motivation, however, in an evolution equation satisfied by
$\omega_{q{\bar q}}$.

Up to corrections that are finite in the
$z\rightarrow 1$ limit, the dilepton hard-scattering function
$\omega_{q{\bar q}}(z,Q^2)$ obeys an evolution equation of the form
\refmark\rgs
$$
\eqalign{
\bigg [ {\partial \over \partial \ln(1/(1-z)) }
&-{1\over 2}\beta(g) {\partial \over \partial g} +1 \bigg ]
\omega_{q{\bar q}}(z,Q^2) \cr
&=
- \int_z^1 d y\,
\bigg [\, { g_1(\alpha_s((1-y)^2Q^2) \over (1-y)} \bigg ]_+
\omega_{q{\bar q}}(y-z,Q^2)
+ g_2(\alpha_s(Q^2))\omega_{q{\bar q}}(z,Q^2)\, , }
\eqn\evoleqn
$$
whose solution is
$$
\eqalign{
\omega_{q{\bar q}}(z,Q^2)
&=
\sum_{n=0}^\infty \prod_{i=0}^n
\int_0^1 d x_i \bigg \{ \bigg [ {1\over 1-x_i} \int_0^x d y_i
{g_1(\alpha_s((1-x_i)(1-y_i)Q^2)) \over 1-y_i} \bigg ]_+
 \cr
 & \hbox{\hskip 1.5 true in} +
 \bigg [ {g_2(\alpha_s((1-x_i)Q^2)) \over (1-x_i)} \bigg ]_+ \bigg \}
\delta(1-z- \sum_{k=1}^n(1-x_k))\, . }
\eqn\evolsoln
$$
Moments of this expression with respect to $n$ yield, up
to corrections of order $1/n$,
$\omega_{q{\bar q}}(n,Q^2)$ in eqs.~\twonine\ and \twoten.
To verify that eq.~\evolsoln\
satisfies \evoleqn, we may convert the derivative of the delta
function with respect to $\ln(1-z)$ into a sum of derivatives
with respect to the $x_k$, and then integrate by parts.  The
remainder of the reasoning only requires the definitions of the
running coupling and the plus distribution.

In terms of the formulation for $\omega_{q{\bar q}}$ just given, we see that
the singularity in the running coupling is already present in
the evolution equation, \evoleqn.  If we choose to
define the $y$ integral of this equation as a principal
value, its solution \evolsoln\ is then given as a sum of
products of principal value $x_i$ integrals, and its moments
become exponentials of principal value integrals.
(The delta functions may be considered as referring to the
real parts of the $x_i$.)
We may,
therefore, wish to consider our principal value prescription
for the exponentiated resummation
as grounded in a
principal value definition of the integro-differential equation
whose solution it is.

Some basic properties of the principal value resummation formula
are immediately obvious.
First, the exponentiated cross section in moment space, eq.~\twoten,  is real
(for real $n$).
This is because the contour $P$
in eq.~\twothirteen\ is a sum of two
mirror-symmetric contours
with respect to the real axis, and the integrand
is real on some portion of the real axis.
Also, $E$ in eq.~\twoten\ is finite by the definition
of $\Gamma_{q{\bar q}}$, eq.~\twotwelve.

At this point, we may come back
to the Borel transformation\refmark\rthree.
All large perturbative corrections\refmark\rtwo are contained in
those pieces of the function
$\Gamma_{q{\bar q}}(1-\zeta, Q^2)$, eq.~\twotwelve,
whose dependence on $\zeta$ and $Q^2$ is of the form
$$\Gamma_{q{\bar q}}(1-\zeta,Q^2)=\sum_{k=0}^2\alpha^k
\Gamma_{q{\bar q}}^{(k)}(\alpha\ln[1/(1-\zeta)])\, ,
\eqn\addone
$$
where we define
$$\
\alpha\equiv\alpha_s(Q^2)/\pi\, ,
\eqn\alphadef
$$
and where each $\Gamma^{(k)}$ is an infinite series in
its arugment.
In our exponent
$$
E(n,Q^2)= -\sum_{k=0}^2\alpha^k
\int_Pd\zeta\biggl({\zeta^{n-1}-1\over 1-\zeta}\biggr)
\Gamma_{q{\bar q}}^{(k)}(\alpha\ln[1/(1-\zeta)])\ ,\eqn\addtwo
$$
we now expand the power of $\zeta$,
$$
\zeta^{n-1}=(1-(1-\zeta))^{n-1}=\sum_{m=0}^\infty (1-n)...(1-n+m-1){1\over m!}
(1-\zeta)^m
=\sum_{m=0}^\infty {(1-n)_m\over m!}(1-\zeta)^m
\eqn\addthree
$$
where `Pochhammer's symbol'
\Ref\reight{M.\ Abramowitz and I.A.\ Stegun, ed.,
{\it Handbook of mathematical functions} (Dover, New York, 1972).}
is defined as $(a)_m\equiv \Gamma(a+m)/\Gamma(a)$.
Performing  the change of variables
$$w=m\alpha\ln[1/(1-\zeta)]\eqn\addfour$$
we arrive at the following form for the exponent:
$$
E(n,Q^2)=-{1\over \alpha}\sum_{k=0}^2\alpha^k\sum_{m=1}^\infty{(1-n)_m\over
m!m}
\gamma_{q{\bar q}}^{(k)}(\alpha;m)\ .
\eqn\addfive
$$
Here, $\gamma_{q{\bar q}}^{(k)}$ is given in
terms of $\Gamma_{q{\bar q}}^{(k)}$, eq.~\addone,  as
$$
\gamma_{q{\bar q}}^{(k)}(\alpha;m)\equiv
\int_{P'}dw{\rm e}^{-w/\alpha}\, \Gamma_{q{\bar q}}^{(k)}(w/m)\ ,
\eqn\addsix
$$
where the principal
value contour $P'$ runs
between $0$ and $\infty$.
This definition of $\gamma_{q{\bar q}}^{(k)}$ exactly coincides
with the
inverse Borel transform,
 defined as a principal value\refmark\rthree.

\section{Evaluation of the Exponent}

We are now ready to discuss the explicit evaluation of the function
$E(n,Q^2)$, eq.~\twothirteen, which is defined in terms of
the functions $g_i$ in
\twotwelve.
The functions $g_i$ are simple expansions in terms of the running coupling
constant
$\alpha_s[\lambda Q^2]$,
with $\lambda$ a scale,
$$
g_i\bigl(\alpha_s[\lambda Q^2]\bigr)=
\sum_{j=1}^\infty \biggl({\alpha_s[\lambda Q^2]\over
\pi}\biggr)^j g_i^{(j)}\ .
\eqn\threeseventeen
$$
The lowest order numerical coefficients, that determine the large perturbative
behavior, are\refmark\rgs\refmark\rfive
$$g_1^{(1)}=2C_F,\ \ g_2^{(1)}=-{3\over 2}C_F,\ \
g_1^{(2)}=C_F\biggl[C_A\biggl(
{67\over 18}-{\pi^2\over 6}\biggr)-{5n_f\over 9}\biggr]\ ,\eqn\addseven$$
where $n_f$ is the number of flavors.
We can relate the resummed expression to one
containing the
fixed coupling $\alpha_s\equiv\alpha_s(Q^2)$, by reexpressing the running
coupling at
 the scaling
variables $(1-\zeta)(1-y)$ and $(1-\zeta)$ in terms of the fixed one, through
the beta
 function.
As remarked in ref.~\rtwo, {\it all} the large perturbative corrections in the
resummation, of order unity or greater
 in the range $\alpha_s\ln[1/(1-z)]<1$, are contained in terms that have
 at least as many
powers of the logarithm of the momentum scale as of the fixed coupling.
This means, in turn, that we only need to keep terms that are leading and
next-to-leading in
the $g_i$ and the QCD beta-function.  We now construct these
terms explicitly.

Consider the renormalization-group equation for the running
coupling
$\alpha\equiv\alpha_s/\pi$,
$$
{\partial\over \partial \ln(\lambda)}\alpha[\lambda Q^2]=
-b_2\bigl(\alpha[\lambda Q^2]
\bigr)^2
-b_3\bigl(\alpha[\lambda Q^2]\bigr)^3\eqn\threeeighteen$$
where, in QCD,
$$b_2=(33-2n_f)/12\ ,\ \ b_3=(306-38n_f)/48\ .\eqn\threenineteen$$
The solution of eq.~\threeeighteen, with the initial condition
$\alpha(\lambda)|_{\lambda=1}=
\alpha\equiv\alpha_s(Q^2)/\pi$,
is
$$
\alpha(\lambda)/\alpha= \bigg [  1+\alpha b_2\ln(\lambda)-(\alpha b_3/b_2)
\big [ \ln(\alpha(\lambda))/\alpha) - \ln((b_2/b_3+\alpha(\lambda))/
(b_2/b_3+\alpha)) \bigg ]^{-1}\ .
\eqn\threetwenty
$$
We can solve this transcendental equation iteratively, keeping only leading
($\alpha^k\ln^k \lambda$)
and next-to-leading ($\alpha^k\ln^{k-1} \lambda$) powers.
These
terms are given by
$$
\alpha(\lambda)/\alpha=
{1\over 1+\alpha b_2 \ln \lambda}-(\alpha b_3/b_2)
{\ln(1+\alpha b_2 \ln \lambda)\over (1+\alpha
 b_2 \ln \lambda)^2}\ .
\eqn\threetwentyone
$$
Combining eqs.~\threeseventeen\ and \threetwentyone\ we find
$$
g_i(\alpha[\lambda Q^2])=\sum_{j=1}^\infty g_i^{(j)}\alpha^j\biggl({1\over
1+\alpha
b_2\ln\lambda}-(\alpha b_3/b_2){\ln(1+\alpha b_2\ln\lambda)
\over (1+\alpha b_2\ln\lambda)^2}\biggr)^j\ .\eqn\threetwentytwo
$$
The leading and next-to-leading terms, which give all large perturbative
corrections,
 will come
from the $j=1,\ 2$ terms only. Keeping just the leading and next-to-leading
pieces in
these terms, we can write $g_i\simeq g_i^L+g_i^{NL}$ with
$$
g_i^L(\alpha[\lambda Q^2])=g_i^{(1)}\alpha{1\over 1+\alpha
b_2\ln\lambda}\eqn\threetwentythree
$$
$$
g_i^{NL}(\alpha[\lambda Q^2])=-g_i^{(1)}\alpha^2(b_3/b_2){\ln(1+\alpha
b_2\ln\lambda)\over
(1+\alpha b_2\ln\lambda)^2}
+g_i^{(2)}\alpha^2{1\over (1+\alpha b_2\ln\lambda)^2}\ .\eqn\threetwentyfour
$$
Note that the distinction between `$L$' and `$NL$' refers only
to the loop order in $g_i$, and not to the powers of logarithms of
$n$ in $E(n,Q^2)$.  In fact, $g_1^{(2)}$, for instance, gives the
same maximum power of logarithms of $n$ as $g_2^{(1)}$.  For now,
however, it will be instructive, and simpler,
 to focus on the one-loop terms only.

Let us now proceed with the analytic evaluation of the exponent
$$
E(n,\alpha)=-\int_P\biggl({\zeta^{n-1}-1\over 1-\zeta}\biggr)\Gamma_{q{\bar
q}}(1-
\zeta,Q^2)
\simeq E(n,\alpha)_L+E(n,\alpha)_{NL}\ .\eqn\threetwentyfive$$
Here and below, we replace $Q^2$ by $\alpha$ as the second
argument of $E$.
$E_L$,
as specified by \twotwelve\ and \threetwentythree, is
$$E(n,\alpha)_L=\alpha(g_1^{(1)}I_1-g_2^{(1)}I_2)\, ,
\eqn\twofourty
$$
with
$$
I_1(t)\equiv t\int_Pd\zeta\biggl({\zeta^{n-1}-1\over
1-\zeta}\biggr)\ln\Biggl({1+(2/t)
\ln(1-\zeta)\over
1+(1/t)\ln(1-\zeta)}\Biggr)=2I(t/2)-I(t)\ ,\eqn\twofourtyonea$$
$$I(t)\equiv t\int_P d\zeta\biggl({\zeta^{n-1}-1\over
1-\zeta}\biggr)\ln(1+(1/t)\ln(1
-\zeta))\, ,
\eqn\twofourtyoneb$$
and
$$
I_2(t)\equiv\int_Pd\zeta\biggl({\zeta^{n-1}-1\over 1-\zeta}\biggr){1\over
1+(1/t)\ln(1
-\zeta)}
\ .
\eqn\twofourtyonec
$$
Here we define
$$
t\equiv 1/(\alpha b_2) = \ln(Q^2/\Lambda^2)\, ,
\eqn\threetwentyeight
$$
where we have used
 the one-loop running coupling, as in eq.~\threetwentythree.

The $\zeta$ integrals in $I_1$ that result from the
expansion \addthree\
are readily carried out by a change of variables to $\ln(1-\zeta)$,
followed by integration by parts, which gives
$$
I_1=\sum_{m=1}^\infty{(1-n)_m\over m!m^2}\{{\cal E}(mt)-2{\cal E} (mt/2)\}\, ,
\eqn\twofiftyfour
$$
where we define the combination
$${\cal E}(x)\equiv x{\rm e}^{-x}Ei(x)\ ,\ x>0\ ,\eqn\threethirtysevenb$$
with the Exponential Integral defined as
the principal value integral\refmark\reight
$$
Ei(x)\equiv {\cal P}\int_{-\infty}^x dy{{\rm e}^y\over y}\ .
\eqn\threethirtysevenc
$$
Similarly we find
$$
I_2=\sum_{m=1}^\infty{(1-n)_m\over m!m}{\cal E}(mt)\ .
\eqn\twofiftyfourb
$$
Hence the leading  exponent can be written as
$$
E(n,\alpha)_L=-\alpha g_1^{(1)}\sum_{m=1}^\infty {(1-n)_m\over
m!m^2}\biggl[2{\cal E}
(mt/2)-{\cal E}(mt)\biggr]
-\alpha g_2^{(1)}\sum_{m=1}^\infty{(1-n)_m\over m!m}{\cal E}(mt)\
.\eqn\twofiftyfourc
$$
Notice in the above formulas that the moment dependence is
entirely contained within the
Pochhammer symbol, and is defined for complex $n$. The
$\alpha_s(Q^2)$-dependence, on the other hand, is contained within the
functions ${\cal E}(mt)$, which are defined for {\it any} value
of $t$ ($\alpha_s$), however small (large).

\section{Higher Twist in the Resummed Exponent}

Eq.~\twofiftyfourc\ affords a direct estimate of the ambiguities
implicit in the pertrubative series.
They are, as expected, of higher twist.
We have treated these ambiguities
by the prinicpal value prescription in eq.~\threethirtysevenc.
Any other prescription for defining the integrals would differ
in the treatment of the singularity at $y=0$.  But the integral
at $y=0$ is proportional to an exponential of $-t$, and is hence
suppressed by a power of $\Lambda/Q$.  The relevant powers may
simply be read off from the arguments of the $\cal E$
in eq.~\twofiftyfourc.  The minimum supression is from ${\cal E}(t/2)$
in the
$m=1$ term, which behaves as $(\Lambda/Q)\ln(Q^2/\Lambda^2)$.  At this
first nonleading power, perturbation theory is already ambiguous,
and nonperturbative effects must come into play.

An alternative to the principal value prescription is to simply cut off
the $\zeta$ integral in eq.~\twotwelve\ at a value large enough to
avoid the singularities of the running coupling.  It is perhaps
worthwile to illustrate the relationship between these two
approaches.
In
ref.~\rfive.
it was shown that the IR cutoff dependence in
 the hard part,
despite being
numerically significant, is higher-twist,
although the precise powers were not determined.
These powers may be easily determined
if we take eq.~\twofiftyfourc\ as a starting point,
redefining the integrals that
define the $\cal E$'s to reflect the cut-off.

For example, consider an
exponent defined by cutting off the
$\zeta$ integral at some minimum
value of $1-\zeta\equiv\xi$. For the integral $I_2$, for example,
the corresponding regulated expression would be
$$I_2^{(reg)}=
\sum_{m=1}^\infty{(1-n)_m\over m!}I_2^{(reg)m}
\eqn\twofiftyfive
$$
with
$$
I_2^{(reg)m}\equiv\int_{\xi_{min}}^1d\xi\xi^{m-1}{1\over 1+(1/t)\ln
\xi}=t\xi_0^m
\int_{m\ln(\xi_{min}/\xi_0)}^{m\ln(1/\xi_0)}dx{{\rm e}^x\over x}\ .
\eqn\twofiftysix
$$
In the second form, $\xi_0=\exp(-t)$ is the position of the pole in $\xi$.
Since $\xi_{min}$ is designed
to make the integral finite, it must be chosen to be larger than the
location of the pole.
To separate only the nonperturbative region, however, we take it to be
the same
 order of magnitude as $\xi_0$.  We parameterize the relation by
$$
 a\equiv\ln(\xi_{min}/\xi_0)\, ,
\eqn\twofiftyseven
$$
with $0<a<1$.
Then, using eqs.~\twofiftyfourb\ and \twofiftysix\ we obtain:
$$
I_2^m-I_2^{(reg)m}=t{\rm e}^{-mt}{\cal P}\int_{-\infty}^{ma}dx{{\rm e}
^x\over x}=t{\rm e}^{-mt}Ei(ma)\ .
\eqn\twofiftyeight
$$
For a given $a$,
the difference between the
cutoff integral and the principal
value integral is exponentially suppressed by the
{\it same} function
of the fixed coupling
constant that appears in the principal value expression
(i.e., by the $m$-th power  of $\Lambda^2/Q^2$).
Exactly analogous results hold for $I_1$.

\chapter{The Asymptotic Series}

Even though eq.~\twofiftyfourc\  gives the result
for the exponent in a form appropriate for numerical
 evaluation,
it is also of interest to study the asymptotic
expansion of $E$ in $\alpha_s(Q^2)$. Notice that the
exponential integrals
in
eq.~\twofiftyfourc\
have a perfectly well-defined Taylor expansion \refmark\reight
$$
Ei(mt)=\gamma+\ln(mt)+\sum_{n=1}^\infty {(mt)^n\over n!n}\, ,
\eqn\Easyexp
$$
which actually converges better than an exponential for any value of $mt$. On
 the other
hand, such a Taylor expansion would reproduce an infinite
 series
of {\it inverse powers} of $\alpha_s$.  The issue
we would
like to address here
is how to recover a perturbative series for the exponent, eq.~\twofiftyfourc.

We can obtain for the special function ${\cal E}(x)$, eq.~\threethirtysevenb,
after repeatedly  integrating by parts, the asymptotic expression
$$
{\cal E}(mt)\simeq\sum_{\rho=0}^{N}{\rho!\over (mt)^\rho}\, ,
\eqn\twosixty
$$
with $N$ chosen to optimize the approximation.
In this asymptotic series, we see explicitly the $\rho !$ behavior at
$\rho$th order, characteristic of infrared renormalons \refmark\rthooft\ .
Their presence is a direct consequence of the singularity in the
perturbative running coupling.

As we shall show in section 4, a common optimum $N$ for all $m$ can
be determined
numerically  for each of the three sums in eq.~\twofiftyfourc. However,
 the
three optimum numbers of asymptotic terms differ for the three sums, so
we will use the (clumsy but we hope clear)
notation  $N\equiv\{N[2I(t/2)],\ N[I(t)],\  N[I_2(t)]\}$
for the three sums.
Using the asymptotic expansion eq.~\twosixty\ in
the expression for $E_L$,
eq.~\twofiftyfourc, the leading exponent becomes a finite perturbative sum:
$$
\eqalign{E(n,\alpha,N)_L & =-\alpha g_1^{(1)}\Bigg \{
\sum_{\rho=0}^{N[2I(t/2)]}\rho !2^{\rho+1}(\alpha b_2)^\rho\sum_{m=1}^\infty
{(1-n)_m\over m!m^{\rho+2}} \cr
& \quad -\sum_{\rho=0}^{N[I(t)]}\rho!(\alpha b_2)^\rho\sum_{m=1}
^\infty{(1-n)_m\over m!m^{\rho+2}}\Bigg\}
-\alpha g_2^{(1)}\sum_{\rho=0}^{N[I_2(t)]} \rho !(\alpha
b_2)^\rho\sum_{m=1}^\infty
{(1-n)_m\over m!m^{\rho+1}}\ .}
\eqn\fourseven
$$
We can reexpress the infinite series in the moment variable in terms of a
plus-distribution, through
the identity
$$\sum_{m=1}^\infty{(1-n)_m\over m!m^{\rho+1}}={(-1)^\rho\over
\Gamma(\rho+1)}\int_0
^1dxx^{n-1}\Biggl(
{\ln^\rho(1-x)\over 1-x}\Biggr)_+\ .\eqn\twosixtytwo$$
These integrals can in turn be expressed, for large $n$, as polynomials in $\ln
n$.
Using the relation
$$\ln^\rho(1-x)=\lim_{\epsilon\to 0^+}\biggl({\partial\over
\partial\epsilon}\biggr)
^\rho (1-x)^\epsilon\, ,
\eqn\foureight
$$
we obtain
$$\sum_{m=1}^\infty{(1-n)_m\over m!m^{\rho+1}}=\lim_{\epsilon\to
0^+}{(-1)^\rho\over
 \Gamma(\rho+1)}\biggl({\partial
\over \partial\epsilon}\biggr)^\rho\biggl\{{\rm B}(n,\epsilon)-{1\over
\epsilon}
\biggr\}\, ,
\eqn\fournine
$$
where B is the Beta function. Using Stirling's formula  for large $n$, we find
$$
{\rm B}(n,\epsilon)\simeq {\rm e}^{-\epsilon\ln n}\Gamma(\epsilon)\
,\eqn\fourten
$$
and
we can approximate the infinite series in the moment variable,
$n$, by a polynomial in $\ln n$:
$$
\sum_{m=1}^\infty {(1-n)_m\over
m!m^{\rho+1}}=(-1)^\rho\sum_{j=0}^{\rho+1}c_{\rho+
1-j}
{(-1)^j\over j!}\ln^jn\ .
\eqn\foureleven
$$
The numbers $c_k$ above are the standard coefficients of the Taylor
expansion\refmark\reight,
$$\Gamma(1+z)=\sum_{k=0}^\infty c_kz^k\, .
\eqn\fourtwelve
$$

To summarize, the leading exponent may be approximated by a
finite perturbative
sum $E(n,\alpha,N)_L$,
$$
E(n,\alpha)_L=E(n,\alpha,N)_L+\Delta(n,\alpha,N)_L\, ,
\eqn\fourfourteen
$$
where the second term on the right is
a remainder which, since it is
not expressible in terms of powers of $\alpha_s$,
contains a higher-twist
contribution. We shall discuss the numerical size of the remainder in section
4.

The truncated expansion $E(n,\alpha,N)_L$ defines
a convergent, resummed perturbative series for the exponent $E$,
and hence for the hard-scattering function.
$N$ is defined in eq.~\fourfourteen, to minimize
$\Delta_L$.
Increasing $N$ further will result in a smaller accuracy (larger $\Delta_L$).
Inversely, for a given accuracy,
there is a {\it maximum} $\alpha_s$ beyond which the approximation of
eq.~\fourfourteen\ breaks down
{\it for any N}, i.e., the remainder is outside the desired accuracy. These
features
of the exponent represent a `nonperturbative barrier' in the accuracy
of the asymptotic approximation, which is
what we might expect.

The leading asymptotic exponent is given by eq.~\foureleven, in \fourseven,
as
$$
\eqalign{E(n,\alpha,N)_L = &\sum_{\rho=1}^{N[2I(t/2)]+1}\alpha^\rho
\sum_{j=0}^{\rho+1}
s^L_{j,\rho}[2I(t/2)]\ln^j n\
+\sum_{\rho=1}^{N[I(t)]+1}\alpha^\rho\sum_{j=0}^{\rho+1}
s^L_{j,\rho}[I(t)]\ln^j n  \cr
& \quad \quad
+\sum_{\rho=1}^{N[I_2(t)+1]}\alpha^\rho\sum_{j=0}^\rho s^L_{j,\rho}[I_2(t)]
\ln^j n }
\eqn\fourfifteen
$$
with
$$
\eqalign{
s^L_{j,\rho}[2I(t/2)] &=-g_1^{(1)}b_2^{\rho-1}(-1)^{\rho+j}{(\rho-1)!
\over j!}2^\rho c_{\rho+1-j}\ ,\cr
s^L_{j\rho}[I(t)] &=g_1^{(1)}b_2^{\rho-1}(-1)^{\rho+j}
{(\rho-1)!\over j!}c_{\rho+1-j}\ ,\cr
s^L_{j,\rho}[I_2(t)] &=g_2^{(1)}b_2^{\rho-1}(-1)^{\rho+j}
{(\rho-1)!\over j!} c_{\rho-j}\ .}
\eqn\foursixteen
$$
Again, the $N$'s
to be kept in the
 asymptotic expansion are
fixed by minimizing the remainder in eq.~\twosixty\
for a fixed value of $\alpha_s(Q^2)$.
We may in principle use
eqs.~\fourfifteen\ and \foursixteen (along with the corresponding results
for $E_{NL}$) to approximate the resummed
normalization by an analytic expression\refmark\rtwo.
Alternately, we may use the full principal-value exponents.
Using either the exact $E$, or its representation as
an asymptotic series,
the
large perturbative corrections\refmark\rtwo to the
hard scattering function in momentum space can be found
from eqs.~\additionb, \additionc.

\chapter{Behavior of the Exponent}

In this section we shall address several
important numerical issues.
Using the `leading' exponent $E(n,\alpha)_L$, we will illustrate the
behavior of the principal value definition and
its approximation as an asymptotic series,
$E(n,\alpha,N)_L$.
We defer a similar discussion for the next-to-leading exponent,
$E(n,\alpha,N')_{NL}$, as well as numerical
results for cross sections, to future work.  We will see
that the principal value exponent behaves in a relatively
mild fashion for all $n$, over a wide range of $Q$, and that
the asymptotic series can give a good approximation to it
unless $n$ is very large.

In the following, we present numerical results for three values of $Q$,
taking $\Lambda=200MeV$.
To be specific, we have used the values
$$\alpha(Q=5GeV)=0.075\ ,\ \ \alpha(Q=10GeV)=0.061\ ,\ \
\alpha(Q=90GeV)=0.039$$
along with\refmark\rtwo
$$b_2(n_f=4)=2.08333\ ,\ \ g_1^{(1)}=8/3\ ,\ \ g_2^{(1)}=-2\ .$$
The optimum $N$, determined numerically, are shown in
table 1.

Several details concerning the numerical calculations
 may be found in the appendix.
Notice from eqs.~\fourfifteen, \foursixteen\ that,
for a given power $\rho$ of the coupling constant,
the maximum power of $\ln n$ in each of the three sums has coefficients
$$\eqalign{s^L_{\rho+1,\rho}[2I(t/2)]&=g_1^{(1)}{b_2^{\rho-1}
\over \rho(\rho+1)}2^\rho\ ,\ \cr
s^L_{\rho+1,\rho}[I(t)]&=-g_1^{(1)}{b_2^{\rho-1}\over \rho(\rho+1)}\ ,\cr
s^L_{\rho,\rho}[I_2(t)]&=g_2^{(1)}{b_2^{\rho-1}\over \rho}\ .}
\eqn\coeff
$$
$$\vbox{\settabs 1\columns
\+\hfill TABLE\ 1\hfill&\cr
\+\hfill Optimum\ numbers\ of\ asymptotic\ terms\ as\ a\ function\ of\ $n$\
and\ $\alpha$
\hfill&\cr}$$
$$\vbox{\settabs 5\columns
\+\hfill$\alpha$\hfill&\hfill$n$\hfill&\hfill$N[2I(t/2)]$
\hfill&\hfill$N[I(t)]$\hfill&\hfill$N[I_2(t)]$\hfill&\cr
\+\hfill0.075\hfill&\hfill20\hfill&\hfill1\hfill&\hfill5\hfill&\hfill5\hfill&\cr
\+\hfill''\hfill &\hfill40\hfill&\hfill1\hfill&\hfill5\hfill&\hfill5\hfill&\cr
\+\hfill''\hfill &\hfill60\hfill&\hfill1\hfill&\hfill5\hfill&\hfill5\hfill&\cr
\+\hfill0.061\hfill&\hfill20\hfill&\hfill2\hfill&\hfill6\hfill&\hfill6\hfill&\cr
\+\hfill''\hfill &\hfill40\hfill&\hfill2\hfill&\hfill6\hfill&\hfill6\hfill&\cr
\+\hfill''\hfill &\hfill60\hfill&\hfill2\hfill&\hfill6\hfill&\hfill7\hfill&\cr
\+\hfill0.039\hfill
&\hfill20\hfill&\hfill4\hfill&\hfill13\hfill&\hfill11\hfill&\cr
\+\hfill''
\hfill&\hfill40\hfill&\hfill4\hfill&\hfill12\hfill&\hfill11\hfill&\cr
\+\hfill''\hfill
&\hfill60\hfill&\hfill4\hfill&\hfill9\hfill&\hfill21\hfill&\cr}$$

{}From table 1 we see that the leading power of $\ln n$ comes by far from
the $N[I(t)]$ and the $N[I_2(t)]$ terms,
 and the corresponding coefficients, as shown in eq.~\coeff\
are both negative.
This shows that the leading exponent, expressed perturbatively as an asymptotic
approximation, becomes negative for sufficiently large values of the moment
$n$.
In momentum space this shows that, within perturbation theory, the
corresponding
cross section tends to a finite limit as $z\to 1$, since the hard part is the
exponential of a quantity that diverges to minus infinity in that region of
phase
space.
Of course, our asymptotic approximation is not valid at the edge of phase
space.
The preceding discussion, however,
 serves to point out that the corresponding resummed
perturbative cross section is finite in that range, and in no need of extra IR
cut-offs.
What is more important, we can reproduce this
satisfactory feature of the hard part without even
restricting ourselves to the perturbative regime. Indeed, we may numerically
calculate
the leading exponent either from eqs.~\twofourty-\twofourtyonec, or from their
analytical
equivalent, eq.~\twofiftyfourc. The calculation
is carried out in moment space,
 but
conclusions can be reached in momentum space with the
correspondence\refmark\rtwo
$$
n\leftrightarrow {1\over 1-z}\ .
\eqn\usualcorr
$$

The calculation of $E(n,\alpha)_L$
from the series in eq.~\twofiftyfourc\
 is pretty sensitive to the accuracy with which intermediate
quantities are calculated, because there are large cancelations involved.
Alternately, as we explain in the appendix,
it is convenient to evaluate the integrals in
eqs.~\twofourtyonea-\twofourtyonec\ numerically
on a deformed
version of the principal value
 contour, which is much more stable from a numerical point of view.
This contour, denoted by ${\bar P}$, is shown in fig. 1.
Some details of the integration may be found in the appendix.
Using these integrals in eq.~\twofourty, we find the solid
curves of fig. 2. On the other hand, we may use the asymptotic expression,
eqs.~\fourfifteen, \foursixteen,
together with values of $N$ like
those in table 1, to obtain a perturbative asymptotic
series
for the leading exponent. The corresponding values of $E(n,\alpha,N)_L$
as a function of $n$ are shown by the dotted curves of fig. 2.

The first thing we note is that the exponent is bounded
at fixed $Q^2$, and
that it reaches its maximum at rather large values of $n$.  Beyond
the maximum, it decreases monotonically toward $-\infty$,
due to the dominance in $E$ of the integral $I_1(t)$, eq.~\twofourtyoneb,
which behaves as $-t\ln(2)\ln(n)$ when $\ln(n)\gg t$.
Comparison between the solid and dotted curves will establish the range
of moments $n$ where the perturbative expression is valid, as a function of
the fixed coupling constant. Below its maximum, $E(n,\alpha)_L$ is nicely
approximated
by its asymptotic series, also shown in fig.~2.
Beyond this, higher twist takes
 over and damps the exponent, eventually
  making it negative. The asymptotic series behaves in an
  analogous manner, but decreases only for much larger $n$.
  In either case, the limit of the exponent is minus infinity
  as $z\rightarrow 1$, and the integral over $z$ in the
  cross section is actually finite for both the full principal
  value exponent and its asymptotic expansion constructed as above.

Introducing the notation
$$n_1\equiv n\biggl(E(n,\alpha)_L={\rm max}\biggr)\ ,\
\ n_2\equiv
n\biggl(E(n,\alpha,N)_L={\rm max}\biggr)$$
and, with $\Delta_L$ given by eq.~\fourfourteen,
$$\delta_L(n,\alpha)\equiv
{|\Delta(n,\alpha,N)_L|\over E(n,\alpha)_L}\times 100\%\ ,$$
we may summarize the main features of this comparison
in the following table:
$$\vbox{\settabs 1\columns
\+\hfill TABLE\ 2\hfill&\cr
\+\hfill The\ exact\ exponent\ versus\ its\ asymptotic\
approximation\hfill&\cr}$$
$$\vbox{\settabs 7 \columns
\+\hfill$\alpha$\hfill&\hfill$n_1$\hfill&\hfill$E(n_1,\alpha)_L$\hfill&
\hfill$E(n_1,\alpha,N)_L$\hfill&\hfill$\delta_L(n_1,\alpha)$\hfill&\hfill$n_2$
\hfill&\hfill$E(n_2,\alpha,N)_L$\hfill&\cr
\+\hfill0.075\hfill&\hfill35\hfill&\hfill1.847\hfill&\hfill1.710\hfill&
\hfill7.4\%\hfill&\hfill680\hfill&\hfill3.833\hfill&\cr
\+\hfill0.061\hfill&\hfill100\hfill&\hfill3.005\hfill&\hfill4.266\hfill&
\hfill41.9\%\hfill&\hfill$98\times 10^3$\hfill&\hfill27.77\hfill&\cr
\+\hfill0.039\hfill&\hfill1880\hfill&\hfill6.635\hfill&\hfill10.954\hfill&
\hfill65.1\%\hfill&\hfill$34\times 10^6$\hfill&\hfill177.5\hfill&\cr}$$

{}From table 2 we conclude that  the asymptotic perturbative
approximation is good nearly up to $n_1$, even though the
error $\delta_L(n_1,\alpha)$ increases significantly with decreasing $\alpha$.
 Beyond $n_1$, the
two curves are numerically very different, showing that the higher-twist
implicit
 in
the solid curves of fig. 2, dominates. Notice also from fig.~2  that, for
$n$ fixed and below $n_1$,
the error decreases (slowly) with decreasing $\alpha$, something
expected from the nature of the asymptotic series. For example,
$\delta_L(35,0.061)=2.6\%$ and $\delta_L(35, 0.039)=2.5\%$.
Finally, we observe from table 2 that, for all three values of $\alpha$ chosen,
$\alpha b_2\ln n_1\simeq 0.6$, while $\alpha b_2\ln n_2\simeq 1-1.4$.

\chapter{Summary and Discussion}

In this paper we have proposed a definition that
removes the perturbative ambiguities  of the QCD resummation
 formula, through a principal value prescription for integrals
 over the running coupling in the
 exponent $E$, defined in
  eq.~\twoten.  We have developed this approach in the
 context of the Drell-Yan cross section, but it should have a more
 general application.

 This resummation procedure provides
an unambiguous (although arbitrary\refmark\rthree)
definition for the resummed
perturbative series for large threshold corrections
in hadron-hadron scattering, to which higher-twist nonperturbative
corrections may, in principle, be added systematically.
We have noted the relation of our principal
value prescription to the method of Borel transformations.

The exponent $E(n,\alpha)$
is analytically calculable, in terms of one
 and two-loop effects that contain all the large perturbative corrections
in the sense discussed in ref.\ 2. For the one-loop
terms of the exponent, these results
 are given as
an infinite series of Exponential Integrals. This
series suggests, in turn, a definition of
the resummed perturbative series, exploiting the asymptotic approximation of
these functions. The
number of terms in the asymptotic exponent is now precisely determinable
numerically,
 as well.

Numerical results show that the exact leading exponent can
enhance
the cross section in a
manner that  is nicely approximated by its asymptotic perturbative
 expansion
up to values of $n$ such that $(\alpha_s(Q^2)/\pi)b_2\ln n\simeq 0.6$.
Beyond this range higher-twist effects
 turn the
exact exponent around to negative values.

We are hopeful that this
approach will be useful in organizing QCD perturbative corrections.
In particular, the results for the exponent shown in fig.\ 2 suggest
that resummed higher-order
radiative corrections may be substantially smaller than the simple
exponentiation of leading-order  results.    This may help to explain the
relative
success of one-loop approximations
at moderate energies.  In addition,
we have seen that nonperturbative effects, whose presence is
implied by ambiguities in resummed perturbation theory, appear
at order $\Lambda/Q$.  For the normalization
of the Drell-Yan cross section then, we may
be in a situation where perturbative
corrections are moderate,
with  relatively large, but kinematically simple, nonperturbative
corrections.  Such a combination could teach us a lot about the
interrelation of these two components of the full theory.

Of course,
 the analytical result for the leading exponent, eq.~\twofiftyfourc, contains
only an (arbitrary) part of the physical higher twist in dilepton
production, and
we have not made a systematic study of higher twist in this paper.
A definition of the
perturbative series in QCD is, however, necessary before higher
twist can be unambiguously defined\refmark\rthree.
Because eq.~\twofiftyfourc\ is
such a well-defined
extension of the perturbative series
for large corrections, it also contains {\it some} higher twist.
This is evident numerically in the
difference
between the functions and their perturbative approximations
 at the edge of phase space in fig.~2.  A complete
 analysis of higher twist in the normalization of the
Drell-Yan cross section is now possible, but must remain the subject
of future work.  One technical observation along these
lines may be worth making here, however.
In addition to the corrections
at order $(\Lambda/Q)^p, p\ge 1$
that we have found above, we also expect corrections of the form
$1/[(1-z)Q^2]$, associated, for instance, with the production of
resonances in the final state.  Such corrections may be equally,
or even more important than those identified here, especially for
very large $Q$.

In future work we hope to explore further the numerical and phenomenological
consequences of this
approach, including numerical results for
 the next-to-leading exponent, and
for the cross section itself.
 This treatment of large perturbative corrections
should also be applicable
to other inclusive hadron-hadron cross sections,
including heavy-quark
 production\Ref\rnine{E.\ Laenen, J.\ Smith and W.L.\ van Neerven, Nucl.\
Phys.\ B369
(1992) 543.} and jet production, where gluon-gluon hard-scattering
functions play an important role.

\ack
We
would like to thank L.\ Alvero, G.\ Levin,
A.\ Mueller, J.\ Smith and W.\ van Neerven for
very helpful discussions.  This work was supported in part by the National
Science Foundation under grant
PHY 9211367 and by the Texas National Research Laboratory.
\endpage

\appendix{Numerical Evaluation of the Exponent}

The first issue we will address is the number of asymptotic terms in the
formula giving $E(n,\alpha_s,N)_L$. As was mentioned in section 3, one
definition could involve a comparison of
each function ${\cal E}(mt)$ with
its asymptotic expansion, eq.~\twosixty. Given that this function is tabulated
and
close to 1 at this range, this would seem a very straightforward approach.
However,
the value of $N$ thus determined would depend on $mt$ and one would have
then to sum that complicated dependence over $m$. We took the simpler
approach of using eq.~\fourseven\  to define
the three numbers $\{N[2I(t/2)],\ N[I(t)],\ N[I_2(t)]\}$ for each of the
individual sums. These numbers of course now
depend on the moment variable $n$, but not strongly. As we show in table 1,
this dependence becomes more  appreciable when the coupling
constant decreases, but this has no numerical consequences for
 the leading asymptotic
exponent in the perturbative regime, precisely because of the smallness of the
corresponding coupling.
The above numbers should be determined
in a range of values of $n$ that is well within the perturbative
restriction $\alpha_s\ln n\ll 1$.
The result is shown in table 1.

Another issue is the numerical evaluation of the exact leading exponent.
If we use the simple and straightforward analytic expression,
eq.~\twofiftyfourc, we observe that the series involved\foot{more accurately,
the
finite sums
involved, if we use integer $n$ (which
we will do in this case),}, although convergent, are alternating. Using double
precision,
one can not reliably calculate the exponent beyond a value $n\simeq 40$,
because
 the alternating partial sums exceed the 16-digit accuracy. Also, the
special functions ${\cal E}(mt)$ should be calculated to a similar accuracy,
and
the corresponding numerical integrations need a lot of computer time.
 This method, for relatively low values of $n$, may be used as a numerical
check for
the  alternative method we mentioned in section 4, namely the use of contour
$\bar P$, fig.~1.
The resulting expressions for the integrals are
rather complicated, from an analytical point of view, but run smoothly on the
computer for much larger values of $n$.
On $\bar P$, the integral $I(t)$, eq.~\twofourtyoneb, can be written as a sum
of
 three
integrals, each one resulting from
twice the real part of the integration along one side of
the contour.
The result is
$$I(t)=I_1(t,r)+I_2(t,r)+I_3(t,r)\eqn\seventwo$$
with $r\equiv n-1$, where:
$$I_1(t,r)=t\int_0^1 {dx\over
x}\biggl(A_1(r;x)B_1(t,r;x)+C_1(r;x)D_1(t,r;x)\biggr)
\eqn\seventhree$$
$$I_2(t,r)=t\int_0^1 {dx\over
x^2+1/r^2}\biggl(A_2(r;x)B_2(t,r;x)+C_2(r;x)D_2(t,r;x)
\biggr)
\eqn\sevenfour$$
$$I_3(t,r)=-t\int_0^{1/r} {dx\over
1+x^2}\biggl(A_3(r;x)B_3(t;x)+C_3(r;x)D_3(t;x)
\biggr)
\eqn\sevenfive$$
and where the various integrands are given by:
$$A_1(r;x)=\biggl(1+x^2/r^2\biggr)^{r/2}\cos(r\arctan(x/r))-1\eqn\sevensix$$
$$B_1(t,r;x)=(1/2)\ln\biggl(\biggl[1+(1/t)\ln(x/r)\biggr]^2+\biggl[\pi/2t\biggr]
^2\biggr)\eqn\sevenseven$$
$$C_1(r;x)=\biggl(1+x^2/r^2\biggr)^{r/2}\sin(r\arctan(x/r))\eqn\seveneight$$
$$D_1(t,r;x)=\arctan\biggl({\pi/2\over t+\ln(x/r)}\biggr) \eqn\sevennine$$
$$\eqalign{A_2(r;x)& =x\biggl(\biggl[(1-x)^2+1/r^2\biggr]^{r/2}
\cos\biggl(r\arctan\biggl({1/r\over 1-x}\biggr)\biggr)-1\biggr)
\cr
&\ \ -(1/r)\biggl[(1-x)^2+1/r^2\biggr]^{r/2}\sin\biggl(r\arctan\biggl({1/r\over
1-x}\biggr)
\biggr)}
\eqn\seventen
$$
$$
B_2(t,r;x)=(1/2)\ln\biggl(\biggl[1+(1/2t)\ln(x^2+1/r^2)\biggr]^2+\biggl[{\arctan
(1/xr)\over t}\biggr]^2\biggr)
\eqn\seveneleven
$$
\endpage
$$
\eqalign{C_2(r;x)&=(1/r)\biggl(\biggl[(1-x)^2+1/r^2\biggr]^{r/2}
\cos\biggl(r\arctan\biggl({1/r)\over 1-x}\biggr)\biggr)-1\biggr)
\cr
&\ \ +x\biggl[(1-x)^2+1/r^2\biggr]^{r/2}\sin\biggl(r\arctan\biggl({1/r\over
1-x}
\biggr)\biggr)}
\eqn\seventwelve
$$
$$D_2(t,r;x)=\arctan\biggl({\arctan(1/rx)\over t+(1/2)\ln(x^2+1/r^2)}\biggr)
\eqn\seventhirteen$$
and
$$A_3(r;x)=x(x^r\cos(r\pi/2)-1)+x^r\sin(r\pi/2)\eqn\sevenfourteen$$
$$B_3(t;x)=(1/2)\ln\biggl(\biggl[1+(1/2t)\ln(1+x^2)\biggr]^2+\biggl[{\arctan
x\over
 t}\biggr]^2
\biggr)\eqn\sevenfifteen$$
$$C_3(r;x)=-(x^r\cos(r\pi/2)-1-x^{r+1}\sin(r\pi/2))\eqn\sevensixteen$$
$$D_3(t;x)=\arctan\biggl({\arctan x\over t+(1/2)\ln(1+x^2)}\biggr)\ .
\eqn\sevenseventeen$$

The integral $I_2$ is handled in a similar fashion.
\endpage
\refout
\endpage
\title {{\bf FIGURE CAPTIONS}}

Figure 1: The contour $\bar P$
used for the numerical evaluation of the leading exponent.
A mirror-symmetric contour in the lower half-plane is
       understood.

Figure 2: (a) The exact
principal value leading exponent (solid curve) versus its asymptotic
approximation (dotted curve) as a function of $n$, for $\alpha(Q=5GeV)\simeq
0.075$.
 (b) Similarly for $\alpha(Q=10GeV)\simeq 0.061$.
 (c) Similarly for $\alpha(Q=90GeV)\simeq 0.039$.
\end